\documentclass[superscriptaddress,showpacs,preprintnumbers,amsmath,amssymb]{revtex4}

\usepackage{graphicx}% Include figure files
\usepackage{dcolumn}% Align table columns on decimal point
\usepackage{bm}% bold math

\begin{document}

\preprint{WU-AP/293/08}

\title{Oscillation and Future Detection of Failed Supernova Neutrinos \\ from Black Hole Forming Collapse}% Force line breaks with \\

\author{Ken'ichiro Nakazato}
 \email{nakazato@heap.phys.waseda.ac.jp}
 \affiliation{Department of Physics, Waseda University, 3-4-1 Okubo, Shinjuku, Tokyo 169-8555, Japan
}%

\author{Kohsuke Sumiyoshi}
\affiliation{Numazu College of Technology, Ooka 3600, Numazu, Shizuoka 410-8501, Japan
}%

\author{Hideyuki Suzuki}
\affiliation{Faculty of Science \& Technology, Tokyo University of Science, Yamazaki 2641, Noda, Chiba 278-8510, Japan
}%

\author{Shoichi Yamada}%
 \altaffiliation[Also at ]{Advanced Research Institute for Science \& Engineering, Waseda University, 3-4-1 Okubo, Shinjuku, Tokyo, Japan} %and Max-Planck-Institut f\"{u}r Astrophysik, Karl-Schwarzshild -Str. 1, D-85741, Garching, Germany.} %Lines break automatically or can be forced with \\
 \affiliation{Department of Physics, Waseda University, 3-4-1 Okubo, Shinjuku, Tokyo 169-8555, Japan
}%

\date{\today}% It is always \today, today,

\begin{abstract}
Recently, stellar collapse involving black hole formation from massive stars is suggested to emit enormous fluxes of neutrinos on par with ordinary core-collapse supernovae. We investigate their detectability for the currently operating neutrino detector, SuperKamiokande. Neutrino oscillation is also taken into account for the evaluation. We find that the event number is larger than or comparable to that of supernova neutrinos and, hence, black hole formation is also a candidate for neutrino astronomy. Moreover, we find that the event number depends dominantly on the equation of state used in the computations of the black hole formation. This fact implies that the detection of neutrinos emitted from the black hole progenitors is very valuable to probe the properties of the equation of state for hot and/or dense matter.
\end{abstract}

%An article usually includes an abstract, a concise summary of the work
%covered at length in the main body of the article. It is used for
%secondary publications and for information retrieval purposes. Valid
%PACS numbers may be entered using the \verb+\pacs{#1}+ command.
%\keywords{}%Use showkeys class option if keyword
%                              %display desired

\maketitle

\section{Introduction} \label{intro}
Detections of neutrinos from SN1987A \cite{hirata87, bionta87} have declared that neutrino astronomy is practical to study high energy astrophysical phenomena. Since future detections of supernova neutrinos will bring useful information not only qualitatively but also quantitatively, there are many studies on the estimations of the event numbers. In particular, about 10,000 events will be detected by SuperKamiokande (SK), which is currently operating neutrino detector, if a supernova occurs near the Galactic center now \cite{kotake06}. It is noted that the neutrino oscillation is discovered \cite{yfukuda99} and confirmed by the solar, atmospheric, reactor and accelerator neutrinos \cite{gongar08}. Since supernova neutrinos propagate through the stellar envelope and the earth, where neutrino flavor conversion occurs by the Mikheyev-Smirnov-Wolfenstein (MSW) effect \cite{wolf78, mik85}, we should take into account it to estimate the event numbers \cite{dighe00, ktaka01, luna01, fogli02, ktaka02, ktaka03a}. There are studies taking into account even the resonant spin-flavor conversion \cite{ando03}, the neutrino self-interaction \cite{duan06} and the effects of the shock propagation \cite{ktaka03b} for supernova neutrinos. The effects of neutrino oscillation on supernova relic neutrino background is also evaluated \cite{ando04}.

It has recently been realized that neutrinos are emitted not only from iron-core collapse supernovae but also from oxygen-neon-magnesium (ONeMg) supernovae \cite{kitaura06}, type Ia supernovae \cite{kunugi07} and black hole forming failed supernovae \cite{sumi06, sumi07, sumi08}. While the flavor conversions of neutrinos from ONeMg supernovae \cite{duan07, luna07} and type Ia supernovae \cite{kunugi07} have been already studied, such a study has not been done of black hole progenitors (hereafter, we call them ``failed supernova neutrinos'') has not yet done.

Stars more massive than $\sim 25$ solar masses ($M_\odot$) will form a black hole at the end of their evolution. Observationally these black hole progenitors have two branches, namely a hypernova branch and a faint supernova branch \cite{nomoto06}. This trend may suggest us that the strongly rotating massive stars result in the hypernova branch while non-rotating massive stars do the faint supernova branch. Stars belonging to the hypernova branch have large kinetic energy and ejected $^{56}$Ni mass and they are implied to associate with the gamma-ray bursts. The numerical studies on the collapse of such stars need fully general relativistic computations for non-spherical situation and they are done merely for the simplified models so far \cite{sekig07}. On the other hand, the kinetic energy and ejected $^{56}$Ni mass are small for stars belonging to the faint supernova branch. The fates of non-rotating massive stars are classified also by a numerical study \cite{fryer99} as follows. a) Stars with main sequence mass, $M\lesssim25M_{\odot}$ make explosions and produce neutron stars. b) Stars ranging $25M_{\odot}\lesssim~M\lesssim40M_{\odot}$ also result in explosions but produce black holes via fallback. c) For $M\gtrsim40M_{\odot}$, the shock produced at the bounce can neither propagate out of the core nor make explosions. Faint supernovae may correspond to class b) of this classification. More recently, a black hole candidate whose mass is estimated as 24-$33M_{\odot}$ is discovered \cite{prest07} and it may be a remnant of class c).

In this study, we focus on the failed supernova neutrinos from class c) objects. In Refs.~\onlinecite{sumi07, sumi08}, collapses of stars with 40 and $50M_{\odot}$ are computed by fully general relativistic neutrino radiation hydrodynamics under spherical symmetry. Having adopted the results of evolutionary calculations for the progenitors \cite{woosley95, hashi95, umeda05, tomi07} as their initial models, they have evaluated the time evolution of emitted neutrino spectra as well as dynamics. As a result, they have concluded that the neutrino luminosity reaches $\sim 10^{53}$ erg/s, which is larger than those of ordinary supernovae \cite{sumi05}. Meanwhile these objects may be observed optically as the disappearance of supergiants because they do not make supernova explosions. In fact, a survey monitoring $\sim10^6$ supergiants is proposed very recently \cite{kocha08}. If one supergiant is confirmed to vanish from this survey, corresponding data of neutrino detectors such like SK should be checked whether the burst of failed supernova neutrinos is detected. The signal of failed supernova neutrinos will provide us various valuable information, for instance, the equation of states (EOS) at high density and temperature \cite{sumi06}. As mentioned already, neutrinos undergo flavor conversion before the arrival at the detectors. One has to evaluate the oscillation effects of failed supernova neutrinos.

We analyze the oscillation and observational aspects of failed supernova neutrinos using the results of Refs.~\onlinecite{sumi07} (hereafter Paper I) and \onlinecite{sumi08} (hereafter Paper II). This paper becomes the first study to deal with them. It is well known that some parameters for the neutrino oscillation are not determined yet. Therefore, we also investigate their dependence. In order to deal the flavor conversion inside the stellar envelope, we utilize the results of evolutionary calculations for the progenitors, which are also adopted in Papers~I and II. When neutrinos pass through the earth before detection, they undergo the flavor conversion also inside the earth. In this case, results of flavor conversions depend also on the nadir angle of the progenitor, which we examine its dependence. The event numbers of failed supernova neutrinos are evaluated for SK.

We arrange this paper as follows. Issues on the neutrino oscillation are given in Sec.~\ref{secosc}. In Sec.~\ref{detect}, we describe the methods to compute the event numbers at SK. The main results are shown in Sec.~\ref{sec:results}. According to Papers~I and II, the feature of neutrino emission differs depending on the EOS and progenitor model. Therefore the resultant neutrino number depends not only on the mixing parameters and nadir angle but also on the EOS and progenitor model. The main purpose of this study is to evaluate their dependence and ambiguities quantitatively. Sec.~\ref{summary} is devoted to a summary.

\section{Neutrino Oscillation} \label{secosc}
In this section, we introduce the general formulation of the flavor conversion by the MSW effect at first. Next, we evaluate the flavor conversion inside the stellar envelope using the results of evolutionary calculations for the stars with $40M_\odot$ \cite{woosley95, hashi95} and $50M_\odot$ \cite{umeda05, tomi07}. Here, a model from Ref.~\onlinecite{woosley95} is adopted as a reference model. Incidentally, this model is an initial condition of the numerical simulations in Paper I. Finally the earth effects are computed from realistic density profile of the earth \cite{dziewo81}.

\subsection{General formulation of flavor conversion by MSW effect}
Here, we review the neutrino oscillation by the MSW effect in brief \cite{kotake06}. Neutrino oscillation is caused by the discrepancy between the mass eigenstate and flavor eigenstate of neutrino. In general, the flavor eigenstate is related with the mass eigenstate as below,
\begin{subequations}
\begin{equation}
|\psi\rangle \equiv \left(
	\begin{array}{ccc}\nu_e\\ \nu_{\mu}\\ \nu_{\tau}
	\end{array}\right)
 = U \left(
	\begin{array}{ccc}\nu_1\\ \nu_2\\ \nu_3
	\end{array}\right),
\label{udef}
\end{equation}
\begin{equation}
U = \left(\begin{array}{ccc}
c_{12}c_{13} & s_{12}c_{13} & s_{13}\\
-s_{12}c_{23}-c_{12}s_{23}s_{13} & c_{12}c_{23}-s_{12}s_{23}s_{13} 
& s_{23}c_{13}\\
s_{12}s_{23}-c_{12}c_{23}s_{13} & -c_{12}s_{23}-s_{12}c_{23}s_{13} 
& c_{23}c_{13}
\end{array}\right),
\label{kobamas}
\end{equation}
\label{evol}
\end{subequations}
where $s_{ij}=\sin\theta_{ij}$, $c_{ij}=\cos\theta_{ij}$ and $\theta_{ij}$ is the mixing angle for $i, j = 1, 2, 3$ ($i<j$). While $\sin^2\theta_{12} \thickapprox 0.32$ and $\sin^2\theta_{23} \thickapprox 0.5$ are well measured from recent experiments, only the upper limit is given for $\sin^2\theta_{13}\leq2.0\times10^{-2}$ \cite{gongar08}. Neutrino propagation is described by the Dirac equation for the mass eigenstate and we can take an extremely relativistic limit here owing to smaller masses of neutrinos comparing with their energy. Moreover, electron type neutrinos get the effective mass from the interaction with electrons when they propagate inside matter. Thus, the time evolution equation of the neutrino wave functions can be written as,
\begin{equation}
i\frac{d}{dt}\left(
	\begin{array}{ccc}\nu_e\\ \nu_{\mu}\\ \nu_{\tau}
	\end{array}\right)
  =   U^{-1} \left(
	\begin{array}{ccc}
		0 & 0 & 0\\
		0 & \Delta m^2_{21} /2E & 0\\
		0 & 0 & \Delta m^2_{31} /2E
	\end{array}\right) U \left(
	\begin{array}{ccc}\nu_e\\ \nu_{\mu}\\ \nu_{\tau}
	\end{array}\right)
  + \left( \begin{array}{ccc}
		\sqrt{2} G_F n_e(t) & 0 & 0\\
		0 & 0 & 0\\
		0 & 0 & 0
	\end{array}\right) \left(
	\begin{array}{ccc}\nu_e\\ \nu_{\mu}\\ \nu_{\tau}
	\end{array}\right),
\label{msw}
\end{equation}
where $G_F$, $n_e(t)$ and $E$ are the Fermi constant, the electron number density and the neutrino energy, respectively. For the anti-neutrino sector, the sign of $n_e(t)$ changes. $t$ is an Affine parameter along the neutrino worldline. $\Delta m^2_{ji}$ are the mass squared differences and they are experimentally determined as $\Delta m^2_{21} = 8\times10^{-5}$~eV and $|\Delta m^2_{31}| = 3\times10^{-3}$~eV. Whether the sign of $\Delta m^2_{31}$ is plus (normal mass hierarchy) or minus (inverted mass hierarchy) is unclear under the current status \cite{gongar08}. Solving Eq.~(\ref{msw}) along the neutrino propagation, we can follow the neutrino flavor eigenstate $|\psi\rangle$ and evaluate the survival probability of $\nu_f$ ($f = e, \mu, \tau$ and their anti-particles) as $p_{\nu_f} = \bigl| \langle \psi_{\nu_f} | \psi \rangle \bigr|^2$.

A calculation of the survival probability can be divided to two parts, namely a probability that a neutrino generated as flavor $f$ reaches to the earth with $i$-th mass eigenstate, $P^\star (\nu_f \to \nu_i)$, and a probability that a neutrino entering the earth with $i$-th mass eigenstate is detected as $\nu_f$, $P^\oplus (\nu_i \to \nu_f)$. Using them, the survival probabilities are expressed as,
\begin{subequations}
\begin{eqnarray}
p_{\nu_e} & = & \sum^{3}_{i=1} P^\star (\nu_e \to \nu_i) P^\oplus (\nu_i \to \nu_e), \\
\label{pee}
p_{\bar{\nu_e}} & = & \sum^{3}_{i=1} P^\star (\bar{\nu_e} \to \bar{\nu_i}) P^\oplus (\bar{\nu_i} \to \bar{\nu_e}), \\
\label{peae}
p_{\nu_x} & \thickapprox & 1 - \sum^{3}_{i=1} P^\star (\nu_\mu \to \nu_i) P^\oplus (\nu_i \to \nu_e), \nonumber \\
 & \thickapprox & 1 - \sum^{3}_{i=1} P^\star (\nu_\tau \to \nu_i) P^\oplus (\nu_i \to \nu_e), \\
\label{pea}
p_{\bar{\nu_x}}  & \thickapprox & 1 - \sum^{3}_{i=1} P^\star (\bar{\nu_\mu} \to \bar{\nu_i}) P^\oplus (\bar{\nu_i} \to \bar{\nu_e}), \nonumber \\
 & \thickapprox & 1 - \sum^{3}_{i=1} P^\star (\bar{\nu_\tau} \to \bar{\nu_i}) P^\oplus (\bar{\nu_i} \to \bar{\nu_e}),
\label{peaa}
\end{eqnarray}
\label{pe}
\end{subequations}
where $\nu_x$ represents the sum of $\nu_\mu$ and $\nu_\tau$, and $\bar{\nu_x}$ represents the sum of $\bar{\nu_\mu}$ and $\bar{\nu_\tau}$. In other words, $p_{\nu_x}$ is a probability that the neutrino generated as $\nu_\mu$ is detected as $\nu_\mu$ or $\nu_\tau$ on the earth. It is noted that we can assume $\mu$-$\tau$ symmetry owing to $\sin^2\theta_{13} \thickapprox 0$ and $\sin^2\theta_{23} \thickapprox 0.5$.

\subsection{Flavor conversion inside stellar envelope}
Neutrinos emitted from the core of black hole progenitor propagate through the stellar envelope, which has the density gradient. In other words, the electron number density, $n_e(t)$, varies along the neutrino worldline. It is noted that the conversions occur mainly in the resonance layers. The matter density at the resonance layer is given as
\begin{equation}
\rho_\mathrm{res.} \sim 6.6 \times 10^5 \left( \frac{\Delta m^2}{1 \, \mathrm{eV}^2} \right) \left( \frac{20 \, \mathrm{MeV}}{E} \right) \left( \frac{0.5}{Y_e} \right) (1-2\sin^2\theta) \, \mathrm{g/cm^3},
\label{res}
\end{equation}
where $Y_e$ is the electron fraction. In the stellar envelope, there are two resonances, namely H resonance and L resonance, where H resonance lies at higher density ($\sim 2\times10^3$~g/cm$^3$) than L resonance does at lower density ($\sim 20$~g/cm$^3$). H resonance corresponds to the case of $\Delta m^2 = \Delta m^2_{31}$ and $\theta = \theta_{13}$ in Eq.~(\ref{res}), while L resonance does to $\Delta m^2 = \Delta m^2_{21}$ and $\theta = \theta_{12}$. Thus, in case of the normal mass hierarchy, both resonances reside for the neutrino sector and the anti-neutrino sector has no resonance. On the other hand, in case of the inverted mass hierarchy, H resonance shifts to the anti-neutrino sector. If the density profile is gradual at the resonance layer, the resonance is ``adiabatic'' and the conversion occurs completely. When the resonance layer has a steep density profile, the resonance is ``non-adiabatic'' and the conversion does not occur. The flip probability of the mass eigenstates, $P_F$, is analytically estimated \cite{bilen87} as
\begin{subequations}
\begin{equation}
P_F = \frac{e^{\chi (1-\sin^2\theta)}-1}{e^\chi-1},
\label{pfl}
\end{equation}
\begin{equation}
\chi \equiv -2\pi \frac{\bigl| \Delta m^2 \bigr|}{2E} \left[ \frac{1}{n_e(r)} \frac{dn_e(r)}{dr} \right]^{-1}_{r=r_\mathrm{res.}},
\label{chi}
\end{equation}
\label{ana}
\end{subequations}
where $r$ and $r_\mathrm{res.}$ are the radial coordinate and the position of the resonance layer, respectively.

Using flip probabilities at H resonance ($P_H$) and L resonance ($P_L$), $P^\star (\nu_f \to \nu_i)$ in Eqs.~(\ref{pe}) are calculated. Here we define matrixes $P^\star$ and $\overline{P^\star}$ whose elements, $P^\star_{if}$ and $\overline{P^\star_{if}}$, are $P^\star (\nu_f \to \nu_i)$ and $P^\star (\bar{\nu_f} \to \bar{\nu_i})$, respectively, and they are calculated as \cite{dighe00}
\begin{subequations}
\begin{eqnarray}
P^\star & = & \left(
	\begin{array}{ccc}
		P_L P_H & 1-P_L & P_L (1-P_H) \\
		(1-P_L)P_H & P_L & (1-P_L) (1-P_H) \\
		1-P_H & 0 & P_H
	\end{array}\right), \\
\label{pstarne}
\overline{P^\star} & = & \left(
	\begin{array}{ccc}
		1 & 0 & 0 \\
		0 & 1 & 0 \\
		0 & 0 & 1
	\end{array}\right),
\label{pstarnae}
\end{eqnarray}
\label{pstarn}
\end{subequations}
in case of the normal mass hierarchy. On the other hand, in case of the inverted mass hierarchy, H resonance resides for the anti-neutrino sector and the survival probabilities are given as,
\begin{subequations}
\begin{eqnarray}
P^\star & = & \left(
	\begin{array}{ccc}
		P_L & 1-P_L & 0 \\
		1-P_L & P_L & 0 \\
		0 & 0 & 1
	\end{array}\right), \\
\label{pstarne}
\overline{P^\star} & = & \left(
	\begin{array}{ccc}
		P_H & 0 & 1-P_H \\
		0 & 1 & 0\\
		1-P_H & 0 & P_H
	\end{array}\right).
\label{pstarnae}
\end{eqnarray}
\label{pstarn}
\end{subequations}

\begin{figure}[t]
\begin{center}
\includegraphics{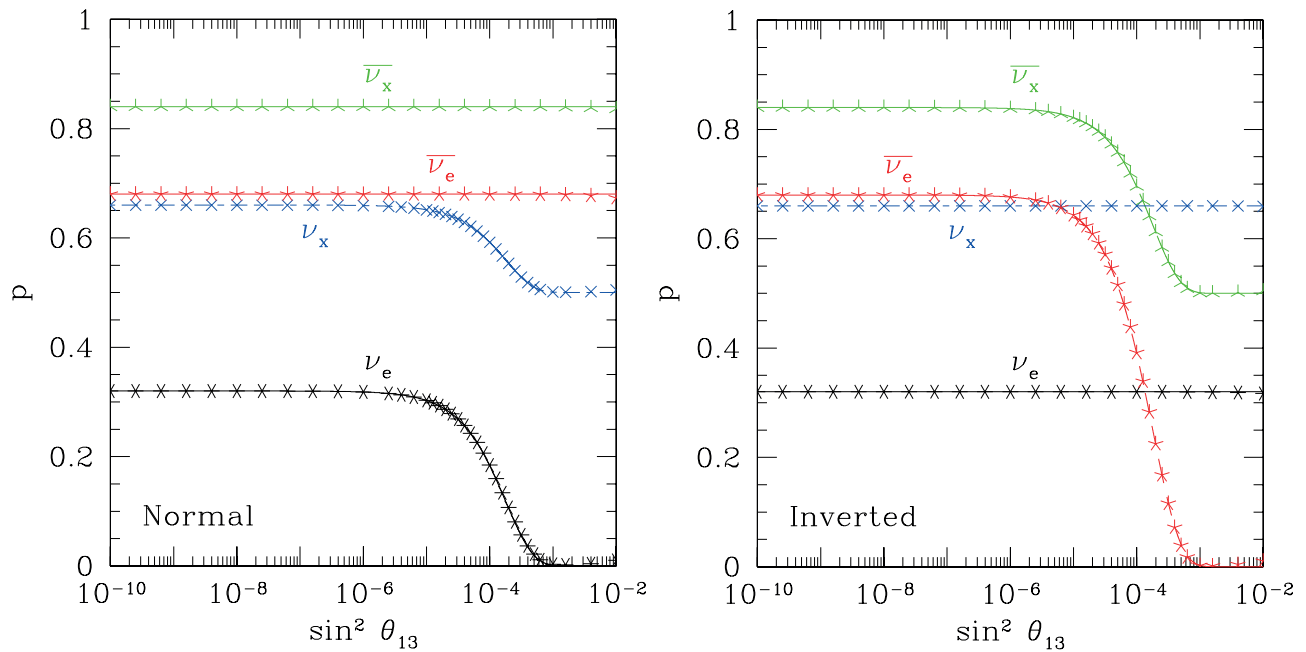}
\caption{Neutrino survival probabilities for the reference model with $E=20$~MeV. Plots represent results of the numerical computations and lines do the analytic formula.}
\label{comp}
\end{center}
\end{figure}

In the absence of the earth effects, $P^\oplus (\nu_i \to \nu_f)$ is displaced to $\bigl| U_{fi} \bigr|^2$, where $U_{fi}$ are elements of matrix $U$ given in Eq.~(\ref{kobamas}), and Eq.~(\ref{pe}) are rewritten as,
\begin{subequations}
\begin{eqnarray}
p_{\nu_e} & \thickapprox & P_H \left[(1-\sin^2\theta_{12})P_L + \sin^2\theta_{12}(1-P_L) \right], \\
\label{pne}
p_{\bar{\nu_e}} & \thickapprox & 1- \sin^2\theta_{12}, \\
\label{pnae}
p_{\nu_x} & \thickapprox & \frac{1+P_H \left[(1-\sin^2\theta_{12})P_L + \sin^2\theta_{12}(1-P_L) \right]}{2}, \\
\label{pna}
p_{\bar{\nu_x}} & \thickapprox & \frac{2- \sin^2\theta_{12}}{2},
\label{pnaa}
\end{eqnarray}
\label{pn}
\end{subequations}
in case of the normal mass hierarchy and
\begin{subequations}
\begin{eqnarray}
p_{\nu_e} & \thickapprox & (1-\sin^2\theta_{12})P_L + \sin^2\theta_{12}(1-P_L), \\
\label{pie}
p_{\bar{\nu_e}} & \thickapprox & P_H (1- \sin^2\theta_{12}), \\
\label{piae}
p_{\nu_x} & \thickapprox & \frac{1+ (1-\sin^2\theta_{12})P_L + \sin^2\theta_{12}(1-P_L)}{2}, \\
\label{pia}
p_{\bar{\nu_x}} & \thickapprox & \frac{1+ P_H(1- \sin^2\theta_{12})}{2},
\label{piaa}
\end{eqnarray}
\label{pi}
\end{subequations}
in case of the inverted mass hierarchy. In Fig.~\ref{comp}, we show the results for the survival probabilities as a function of $\sin^2\theta_{13}$ for the reference model with $E=20$~MeV. In this figure, the analytic formula, Eqs.~(\ref{pn}) and (\ref{pi}), are compared with the numerical solutions of Eq.~(\ref{msw}) by Runge-Kutta method. We can recognize that the analytic estimations well coincide with numerical results. This trend holds true for other progenitor models and neutrino energies, $E$, and we can safely use the analytic formula hereafter. In addition, L resonance is perfectly adiabatic (i.e., $P_L\thickapprox0$) for all of the progenitor models adopted here and for an energy range of our interest ($E=1$-250~MeV). Thus, in case of the normal mass hierarchy $p_{\bar{\nu_e}}\thickapprox0.68$ and $p_{\bar{\nu_x}}\thickapprox0.84$ are robust while $p_{\nu_e}\thickapprox0.32$ and $p_{\nu_x}\thickapprox0.66$ are in the inverted mass hierarchy.

\subsection{Progenitor dependence}\label{envprogen}
\begin{figure}[t]
\begin{center}
\includegraphics{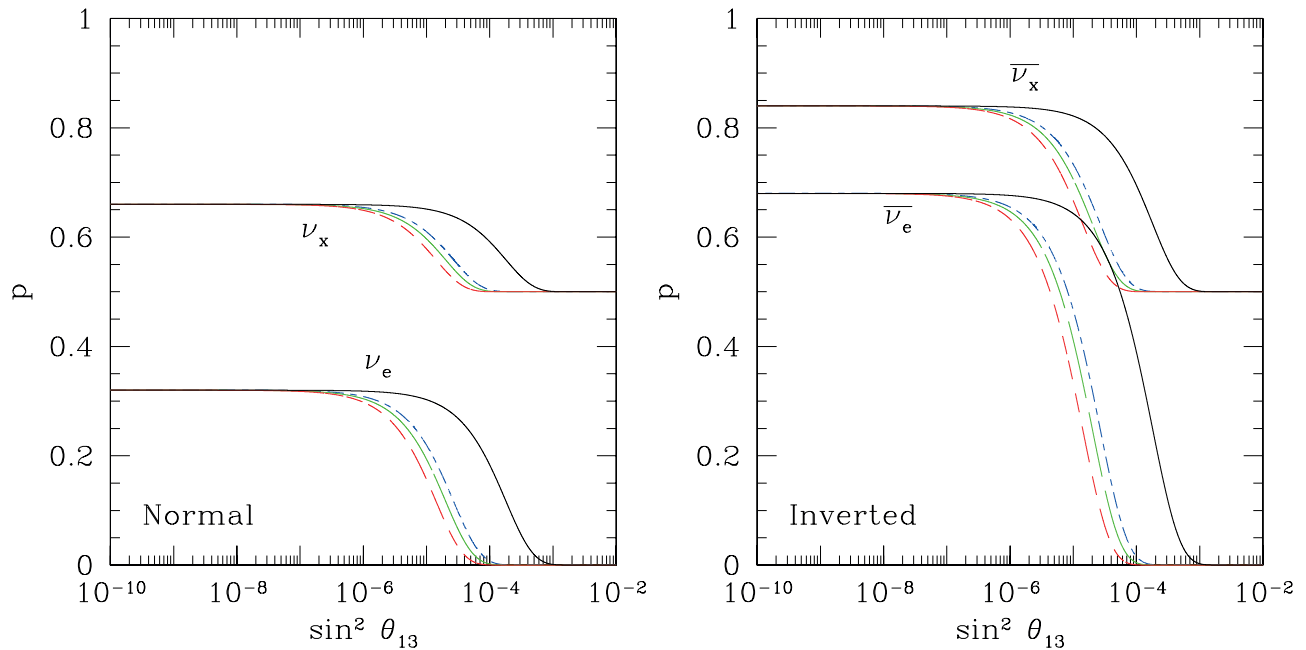}
\caption{Survival probabilities of neutrinos with $E=20$~MeV. Solid, dot-dashed, short-dashed and long-dashed lines represent models by WW95 \cite{woosley95}, H95 \cite{hashi95}, TUN07 \cite{umeda05, tomi07} and TUN07Z \cite{umeda07}, respectively.}
\label{prog}
\end{center}
\end{figure}
In Paper II, the progenitor dependence of neutrino emission is investigated. Since the neutrino survival probabilities depend on the density profile of stellar envelope, we utilize the same stellar models of evolutionary calculations with Paper II. $40M_\odot$ model by Ref.~\onlinecite{woosley95} (the reference model), $40M_\odot$ model by Ref.~\onlinecite{hashi95} and $50M_\odot$ model by Refs.~\onlinecite{umeda05, tomi07} are named WW95, H95 and TUN07, respectively, in Paper II. We use the same notation in our paper. Incidentally, these models do not take into account the mass-loss whereas massive stars are suggested to lose their mass during the quasi-static evolutions. If strong a stellar wind sheds a large amount of the envelope, the density profile at the resonance layer will be affected. The mass-loss rate is thought to depend on the stellar metalicity, $Z$. Since TUN07 is a model with $Z=0$, the effects of mass-loss are absent. Incidentally the same model with TUN07 but $Z=0.02$ (solar metalicity) has also been computed \cite{umeda07}. In this paper, we adopt this model as TUN07Z in order to discuss its sensitivity. In addition, WW95 and H95 are solar metalicity models but without the mass loss.

In Fig.~\ref{prog}, we show $p_{\nu_e}$ and $p_{\nu_x}$ for the normal mass hierarchy and $p_{\bar{\nu_e}}$ and $p_{\bar{\nu_x}}$ for the inverted mass hierarchy as functions of $\sin^2\theta_{13}$ evaluated from the analytic formula. The neutrino energy is fixed to $E=20$~MeV for all models. We can see that the results of model WW95 differs from others. This is because the H resonance lies at the boundary of oxygen layer and carbon layer for model WW95 and its density gradient is steeper than those of other models. 

\begin{figure}[t]
\begin{center}
\includegraphics{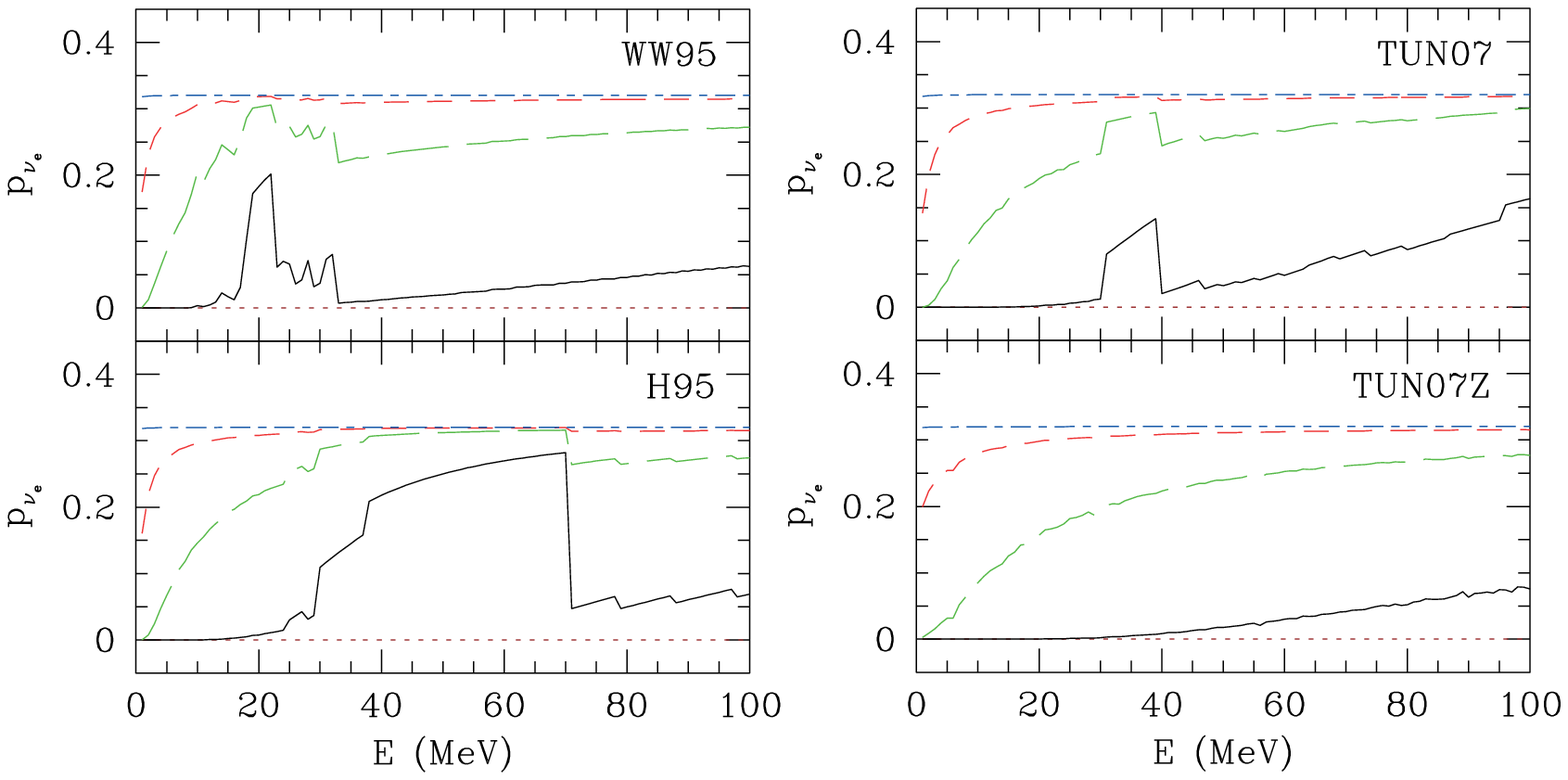}
\caption{Energy dependence of the survival probabilities of $\nu_e$ for each progenitor model. Dotted, solid, long-dashed, short-dashed and dot-dashed lines represent the cases of $\sin^2\theta_{13}=10^{-2}$, $10^{-4}$, $10^{-5}$, $10^{-6}$ and $10^{-8}$, respectively. The normal mass hierarchy is posited here.}
\label{ener}
\end{center}
\end{figure}

As already seen in Eq.~(\ref{res}), the resonance layer depends on the neutrino energy, $E$. In Fig.~\ref{ener}, the energy dependences of $p_{\nu_e}$ for each progenitor model are shown for various values of $\sin^2\theta_{13}$ in case of the normal mass hierarchy. $p_{\nu_e}$ is insensitive to the progenitor model for $\sin^2\theta_{13}\geq10^{-3}$ or $\sin^2\theta_{13}\leq10^{-6}$ whereas it is sensitive for $10^{-6}\leq\sin^2\theta_{13}\leq10^{-3}$. In case of intermediate value of $\sin^2\theta_{13}$, there are energies which give a peak of $p_{\nu_e}$ for some models. Roughly speaking, H resonance of neutrinos with this energy lies at the boundary of layers. For instance, model TUN07 has the boundary of convective region and non-convective region near He shell burning layer. On the other hand, the envelope of model TUN07Z is already stripped and there are no such a boundary near the resonance layer. Thus, its result does not have a peak of $p_{\nu_e}$ even for the case of intermediate value of $\sin^2\theta_{13}$. These trends hold true for $p_{\nu_x}$ in case of the normal mass hierarchy, and for $p_{\bar{\nu_e}}$ and $p_{\bar{\nu_x}}$ in case of the inverted mass hierarchy.

\subsection{Earth effects: nadir angular dependence}\label{naddep}
\begin{figure}[t]
\begin{center}
\includegraphics{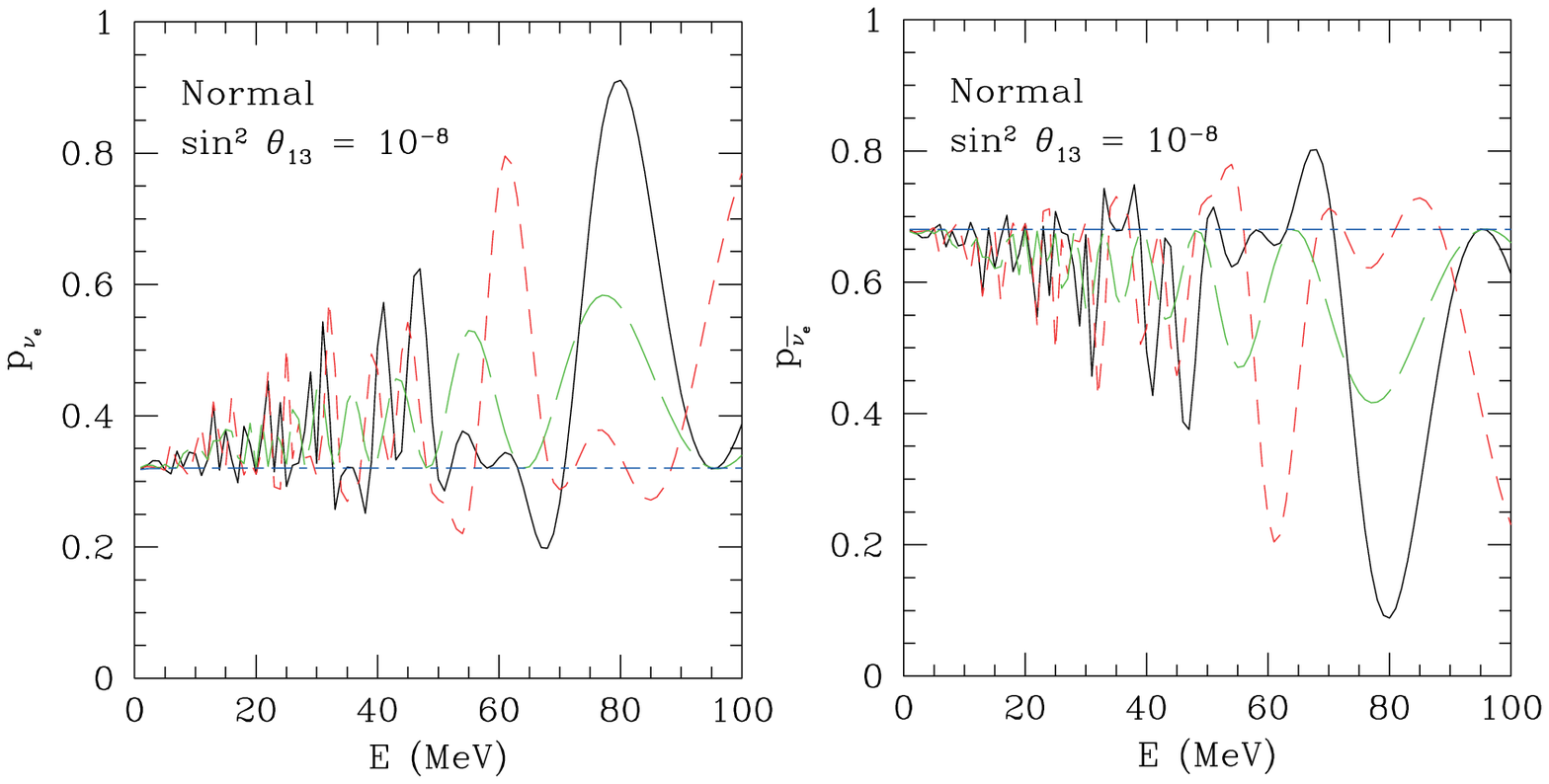}
\caption{Energy dependence of the neutrino survival probabilities with the earth effects for the reference model. Solid, short-dashed, long-dashed and dot-dashed lines represent the cases of the nadir angle $\theta_\mathrm{nad}=0^\circ$, $30^\circ$, $60^\circ$ and $90^\circ$, respectively. Here $\theta_\mathrm{nad}=90^\circ$ is equivalent to the case without the earth effects.}
\label{nadir}
\end{center}
\end{figure}

In this paper, we evaluate the earth effects, $P^\oplus (\nu_i \to \nu_f)$, solving Eq.~(\ref{msw}) along the neutrino worldline numerically by Runge-Kutta method using the realistic density profile of the earth \cite{dziewo81}. It is noted that the density variation along the neutrino worldline depends the nadir angle of the black hole progenitor. In Fig.~\ref{nadir}, we show some examples for the results of the earth effects. We find that the spectral shape is deformed to wave-like and its shape varies with the nadir angle. This is because the typical length of neutrino oscillation becomes comparable to the size of the earth and the probabilities $P^\oplus (\nu_i \to \nu_f)$ become sensitive to the neutrino energy. The deformation of the spectrum occurs for most of the cases of the parameter sets. However, when the value of $\sin^2\theta_{13}$ is larger ($\gtrsim 10^{-3}$) and H resonance is perfectly adiabatic, the deformation disappears for the neutrino sector in case of the normal mass hierarchy and for the anti-neutrino sector in case of the inverted mass hierarchy. Incidentally, these features of deformation are very similar with the case of the ordinary supernovae \cite{luna01, ktaka02} and ONeMg supernovae \cite{luna07}.

\section{Setups for Failed Supernova  Neutrino Detection}\label{detect}

\subsection{Neutrino flux and spectrum before oscillation}\label{sumimodel}
In this study, results in Paper I and II are adopted as the number spectra of the failed supernova neutrinos before the oscillation, $dN_{\nu_f}/dE$. The numerical code of general relativistic $\nu$-radiation hydrodynamics adopted in these references solves the Boltzmann equation for neutrinos together with Lagrangian hydrodynamics under spherical symmetry. Four species of neutrinos, namely $\nu_e$, $\bar\nu_e$, $\nu_\mu$ and $\bar\nu_\mu$, are considered assuming that $\nu_\tau$ and $\bar\nu_\tau$ are the same as $\nu_\mu$ and $\bar\nu_\mu$, respectively. It is noted that the luminosities and spectra of $\nu_\mu$ and $\bar\nu_\mu$ at the emission (before the oscillation) are almost identical because they have the same reactions and the difference of coupling constants is minor. 14 grid points are used for energy distribution of neutrinos. We interpolate the energy spectra linearly, conserving the total neutrino number. We calculate the probabilities of flavor conversion and event numbers at the detector. Taking into account the oscillation, spectra of neutrino flux at the detector, $dF_{\nu_f}/dE$, can be expressed as:
\begin{subequations}
\begin{eqnarray}
\frac{dF_{\nu_e}}{dE} & = & \frac{1}{4 \pi R^2} \left[ p_{\nu_e} \frac{dN_{\nu_e}}{dE} + (1-p_{\nu_x}) \frac{dN_{\nu_x}}{dE} \right], \\
\label{fle}
\frac{dF_{\bar{\nu_e}}}{dE} & = & \frac{1}{4 \pi R^2} \left[p_{\bar{\nu_e}} \frac{dN_{\bar{\nu_e}}}{dE} + (1-p_{\bar{\nu_x}}) \frac{dN_{\bar{\nu_x}}}{dE} \right], \\
\label{flae}
\frac{dF_{\nu_x}}{dE} & = & \frac{1}{4 \pi R^2} \left[ (1-p_{\nu_e}) \frac{dN_{\nu_e}}{dE} + p_{\nu_x} \frac{dN_{\nu_x}}{dE} \right],  \\
\label{fla}
\frac{dF_{\bar{\nu_x}}}{dE} & = & \frac{1}{4 \pi R^2} \left[(1-p_{\bar{\nu_e}}) \frac{dN_{\bar{\nu_e}}}{dE} + p_{\bar{\nu_x}} \frac{dN_{\bar{\nu_x}}}{dE} \right], 
\label{flaa}
\end{eqnarray}
\label{fl}
\end{subequations}
where $R$ is a distance from a black hole progenitor to the earth. In this study, we set $R=10$~kpc, which is typical length of our Galaxy.

\begin{figure}[t]
\begin{center}
\includegraphics{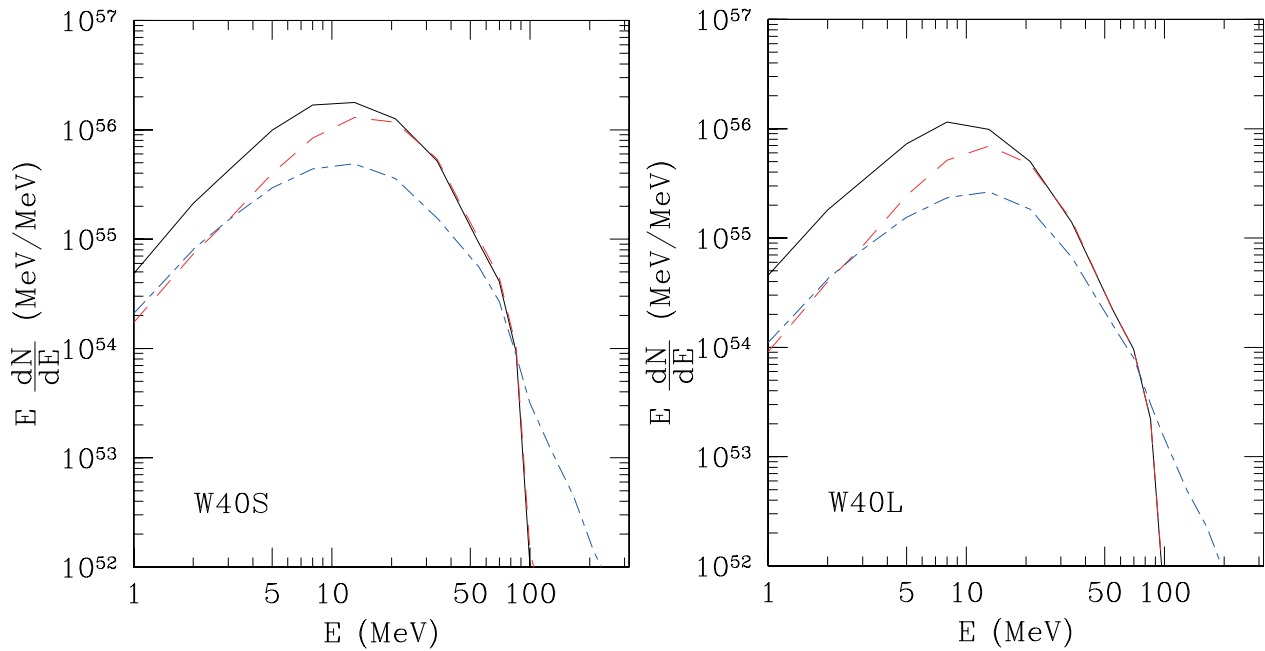}
\caption{Time-integrated spectra before the neutrino oscillation for models~W40S (left) and W40L (right). Solid, dashed and dot-dashed lines represent the spectra of $\nu_e$, $\bar\nu_e$ and $\nu_x$, respectively.}
\label{inispec}
\end{center}
\end{figure}

In Papers I and II, the core collapse of massive stars with 40 and $50M_\odot$~\cite{woosley95, hashi95, umeda05, tomi07} have been computed using two sets of supernova EOS's. They have concluded that the EOS at high density and temperature can be constructed from the duration of failed supernova neutrinos. In these references, they have adopted the EOS by Lattimer \& Swesty (1991) (LS-EOS) \cite{lati91} and EOS by Shen et~al. (1998) (Shen-EOS) \cite{shen98a, shen98b}. LS-EOS is obtained by the extension of the compressible liquid model with three choices of the incompressibility. The case of 180~MeV, which is a representative of soft EOS, has been chosen in Paper I and II. Shen-EOS is based on the relativistic mean field theory and a rather stiff EOS with the incompressibility of 281~MeV. Progenitor models adopted in Paper I and II are already stated in Sec.~\ref{envprogen} and their calculated models are listed in Table~\ref{sumilist}. We use the same names for the models given in Paper I and II hereafter. Since Shen-EOS is stiffer than LS-EOS, the maximum mass of the neutron star is larger and it takes longer time to form a black hole. Therefore the total energy of emitted neutrinos is larger for Shen-EOS than LS-EOS. In Fig.~\ref{inispec}, time-integrated spectra without the neutrino oscillation are shown for models~W40S and W40L. The plots for $\bar\nu_x$ are not shown because they are almost the same with those for $\nu_x$. However, this feature is not the case after the oscillation is taken into account.
\begin{table}[t]
\caption{Summary of calculated models in Paper I and II. For the names of progenitors and EOS's, see in Sec.~\ref{envprogen} and Sec.~\ref{sumimodel}, respectively.  $M_\mathrm{prog}$ and $t_\mathrm{BH}$ are the mass of progenitor and the time at the black hole formation measured from the core bounce, respectively. $E^\mathrm{tot}_\nu$ is the total energy of emitted neutrinos for all species. The averaged neutrino energy of all species is defined as $\langle E_\mathrm{all} \rangle \equiv E^\mathrm{tot}_\nu / N^\mathrm{tot}_\nu$, where $N^\mathrm{tot}_\nu$ is the total number of emitted neutrinos for all species.}
\begin{center}
\setlength{\tabcolsep}{8pt}
\begin{tabular}{ccccccc}
\hline\hline
 Model & Progenitor & $M_\mathrm{prog}$ ($M_\odot$) & EOS & $t_\mathrm{BH}$ (s) & $E^\mathrm{tot}_\nu$ (ergs) & $\langle E_\mathrm{all} \rangle$ (MeV) \\ \hline
 W40S & WW95 & 40 & Shen & 1.35 & $5.15\times10^{53}$ & 23.6 \\
 W40L & WW95 & 40 & LS & 0.57 & $2.03\times10^{53}$ & 20.2 \\
 T50S & TUN07 & 50 & Shen & 1.51 & $4.94\times10^{53}$ & 23.7 \\
 T50L & TUN07 & 50 & LS & 0.51 & $1.66\times10^{53}$ & 19.7 \\
 H40L & H95 & 40 & LS & 0.36 & $1.31\times10^{53}$ & 18.6 \\
\hline\hline
\end{tabular}
\label{sumilist}
\end{center}
\end{table}

\subsection{Neutrino events at SuperKamiokande}
SuperKamiokande is a water Cherenkov detector, which is a descendant of Kamiokande, located at the site of Kamioka mine in Japan. The neutrino reactions in the detector which we take into account are as follows:
\begin{subequations}
\begin{eqnarray}
\label{ibd}
\bar\nu_e + p & \longrightarrow & e^+ + n, \\
\label{nes}
\nu_e + e & \longrightarrow & \nu_e + e, \\
\label{anes}
\bar\nu_e + e & \longrightarrow & \bar\nu_e + e, \\
\label{nas}
\nu_x + e & \longrightarrow & \nu_x + e, \\
\label{anas}
\bar\nu_x + e & \longrightarrow & \bar\nu_x + e, \\
\label{ibdf}
\nu_e + \,^{16}\mathrm{O} & \longrightarrow & e + \,^{16}\mathrm{F}, \\
\label{ibdn}
\bar\nu_e + \,^{16}\mathrm{O} & \longrightarrow & e^+ + \,^{16}\mathrm{N},
\end{eqnarray}
\label{sk3}
\end{subequations}
We adopt the cross sections for the reaction (\ref{ibd}) from Ref.~\onlinecite{strumia03}, for the reactions (\ref{ibdf}) and (\ref{ibdn}) from Ref.~\onlinecite{haxton87} and for others from Ref.~\onlinecite{totsuka92}. It is noted that the reaction (\ref{ibd}) makes dominant contribution to the total event number. Incidentally, the reaction (\ref{ibdf}) has the largest cross section for neutrinos with the energy $\gtrsim 80$~MeV although the ambiguities exist the theoretical cross sections of the neutrino-nucleus reactions. We assume that the fiducial volume is 22.5~kton and the trigger efficiency is 100\% at 4.5~MeV and 0\% at 2.9~MeV, which are the values at the end of SuperKamiokande I \cite{hosaka06}. The energy resolution was 14.2\% for $E_e=10$~MeV at this phase \cite{hosaka06} and roughly proportional to $\sqrt{E_e}$ \cite{luna01}, where $E_e$ is the observed kinetic energy of scattered electrons and positrons. In this study, we choose an energy bin width of 1~MeV.

\section{Results and Discussions}\label{sec:results}
In this section, we report the time-integrated event number of failed supernova neutrinos at SK in the setup discussed above. The flavor conversions of neutrinos are computed from the density profiles of corresponding progenitors in Sec.~\ref{envprogen} for 5 models listed in Table~\ref{sumilist}. We investigate 2 additional models, Tz50S and Tz50L, whose models of emitted neutrinos are the same as T50S and T50L, respectively, but neutrino oscillations are calculated under the density profile of model TUN07Z. We note again that the progenitor model TUN07Z is the same as TUN07 but with the effects of the mass-loss. In Sec.~\ref{paradep}, dependence on the undetermined parameters, namely $\sin\theta_{13}$ and the mass hierarchy, and the nadir angle $\theta_\mathrm{nad}$ is investigated. Dependence on the EOS's and progenitor models is shown in Sec.~\ref{modeldep}. Quantitative discussions on how these ambiguities affect the results are will be also made.

\subsection{Parameter dependence}\label{paradep}
\begin{figure}[t]
\begin{center}
\includegraphics{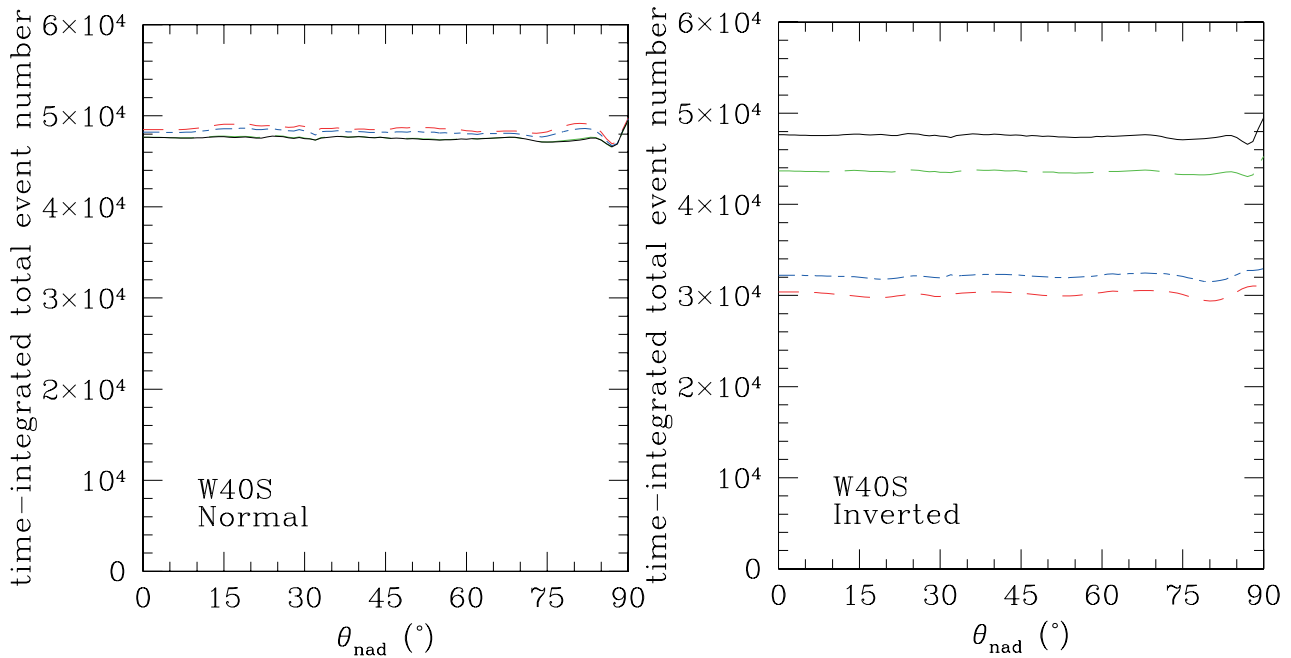}
\caption{Nadir angular dependence of the time-integrated total event number of failed supernova neutrinos for the model~W40S. Solid, long-dashed, dot-dashed and short-dashed lines represent the cases of $\sin^2\theta_{13}=10^{-8}$, $10^{-5}$, $10^{-4}$ and $10^{-2}$, respectively.}
\label{paradepf}
\end{center}
\end{figure}

Here we focus on model~W40S only since the qualitative features are similar among other models. In Fig.~\ref{paradepf}, we show the time-integrated event number of neutrinos for several $\sin^2\theta_{13}$ values as a function of the nadir angle. The event number does not depend on the nadir angle very much and we will revisit later for the reason. It is also noted that the total event number is larger than that of the ordinary supernova ($\sim10,000$) \cite{kotake06} because the total and averaged energies of emitted neutrinos are higher for the black hole progenitors. For instance, for the ordinary supernovae, the total and averaged energies of the neutrinos emitted as $\bar{\nu_e}$ are $4.7 \times 10^{52}$~ergs and 15.3~MeV, respectively~\cite{totani98}, while they are $1.42 \times 10^{53}$~ergs and 24.4~MeV in our model. Even the sum of total energies of emitted neutrinos over all species is lower in the ordinary supernova as $2.9 \times 10^{53}$~ergs.

In case of the inverted mass hierarchy, the event number gets smaller for the larger $\sin^2\theta_{13}$. This is because the survival probability of $\bar\nu_e$, which contributes to the event number most by the reaction (\ref{ibd}), is almost zero for $\sin^2\theta_{13} \gtrsim 10^{-3}$ (Fig.~\ref{prog}). It is important to note that this tendency is different from the case of ordinary supernovae. For the ordinary supernova neutrinos, the flux of $\bar\nu_x$ before the oscillation is larger than that of $\bar\nu_e$ in the energy regime $E \gtrsim 25$~MeV and the neutrino flux is not so small for this regime (see Fig.~35 in Ref.~\onlinecite{kotake06}). For the case of inverted mass hierarchy and larger $\sin^2\theta_{13}$, as seen in Fig.~\ref{prog}, half of $\bar\nu_x$ converts to $\bar\nu_e$ and the event number of neutrinos with $E \gtrsim 25$~MeV increases. Thus the resultant total event number of ordinary supernova neutrinos becomes larger. On the other hand, for failed supernova neutrinos, the neutrino flux is very small for the energy regime ($E \gtrsim 80$~MeV) where the flux of $\bar\nu_x$ before the oscillation becomes larger than that of $\bar\nu_e$ (see Fig.~\ref{inispec}).

The event number gets larger for the larger $\sin^2\theta_{13}$ in the normal mass hierarchy while the difference is small. The difference owing to $\sin^2\theta_{13}$ is mainly induced by the reaction (\ref{ibdf}). From Fig.~\ref{inispec}, we can see that the differential number of $\nu_x$ before the oscillation is larger than that of $\nu_e$ for $E \gtrsim 80$~MeV. Thus the differential number of $\nu_e$ becomes larger by the oscillation for $E \gtrsim 80$~MeV while it does smaller for $E \lesssim 80$~MeV. Incidentally, the threshold of this reaction is high $\sim 15.4$~MeV and the cross section is roughly proportional to $E^2$. Therefore the difference arises crucially for the high energy regime and resultant total event number is slightly larger for the larger $\sin^2\theta_{13}$. It should be remarked, however, that the precise evaluation for the high energy tail of neutrino spectra is difficult in the numerical computations and our results are thought to have some quantitative ambiguities. In addition, this feature is consistent with the case of ordinary supernovae. It is also noted that the mass hierarchy does not make any difference for $\sin^2\theta_{13} \lesssim 10^{-6}$, as is the case of ordinary supernovae.
\begin{figure}
\begin{center}
\includegraphics{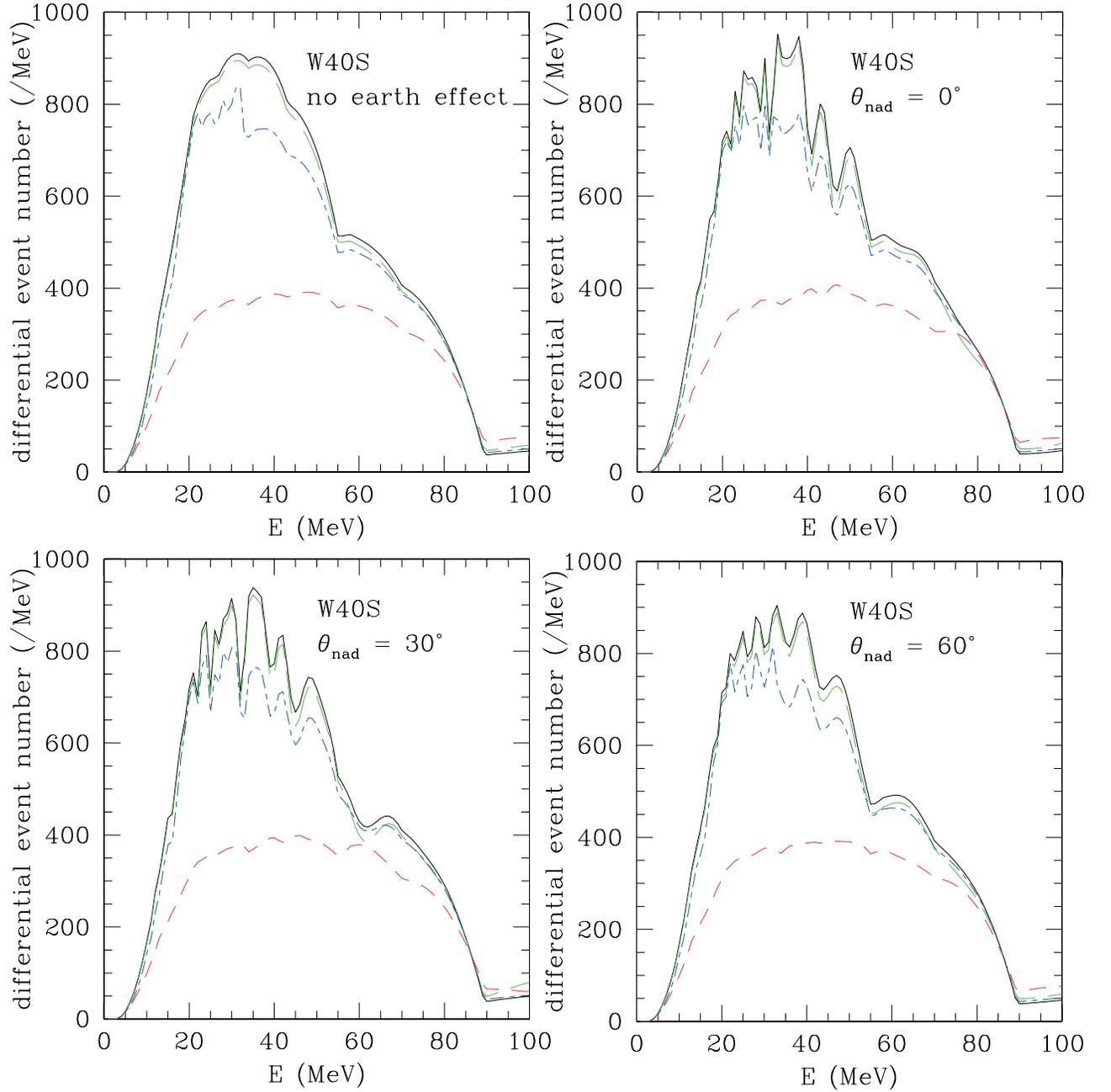}
\caption{Energy spectra for the time-integrated event number of failed supernova neutrinos for the model~W40S in the cases without the earth effects (upper left) and the nadir angle $\theta_\mathrm{nad}=0^\circ$ (upper right), $30^\circ$ (lower left) and $60^\circ$ (lower right). Solid and long-dashed lines represent the cases of $\sin^2\theta_{13}=10^{-8}$ and $10^{-2}$, respectively, for the normal mass hierarchy. Dot-dashed and short-dashed lines represent the cases of $\sin^2\theta_{13}=10^{-5}$ and $10^{-2}$, respectively, for the inverted mass hierarchy.}
\label{paradeps}
\end{center}
\end{figure}

In Fig.~\ref{paradeps}, we show the energy spectra of the time-integrated event number for several models. Results of the models with $\sin^2\theta_{13}=10^{-5}$ for the normal mass hierarchy and $\sin^2\theta_{13}=10^{-8}$ for the inverted mass hierarchy are not shown because they are very close to the result of the model with $\sin^2\theta_{13}=10^{-8}$ for the normal mass hierarchy. The most outstanding case is the inverted mass hierarchy with $\sin^2\theta_{13}=10^{-2}$. The event number is much smaller than those of the other cases especially for the energy regime, $E\lesssim50$~MeV, and accordingly the spectrum becomes harder. The reason is again that the survival probability of $\bar\nu_e$ is almost zero and the original flux of $\bar\nu_e$ is larger than that of $\bar\nu_x$ especially for this energy regime. This spectral feature may be useful for the restrictions of the mixing parameters, as already studied for ordinary supernovae \cite{kotake06}.

In the case without the earth effects, spectra are not fluctuating while the cusps are artifacts due to the binning of original spectral data in Paper I and II. One exception is the model with $\sin^2\theta_{13}=10^{-5}$ for the inverted mass hierarchy, whose spectrum is fluctuating from 20~MeV to 30~MeV. This is because, as already mentioned in Sec.~\ref{envprogen}, the H resonance lies at the boundary of shells for the progenitor model~WW95. The fluctuation originates from spikes in the upper left panel of Fig.~\ref{ener}. In the case with the earth effects for the nadir angle $\theta_\mathrm{nad}=0^\circ$, $30^\circ$ or $60^\circ$, fluctuating spectra are from the earth effects. This is because the spectral shapes of survival probabilities for each neutrino species are deformed to wave-like as shown in Sec.~\ref{naddep}. In the case of the inverted mass hierarchy with $\sin^2\theta_{13}=10^{-2}$, where there is no earth effect on the anti-neutrino sector, the deformation of spectra is not clear comparing with the other cases. Integrating over the neutrino energy, the fluctuation is smoothed out. Therefore the total event number does not depend on the nadir angle very much.

\subsection{EOS and stellar model dependence}\label{modeldep}
\begin{figure}[t]
\begin{center}
\includegraphics{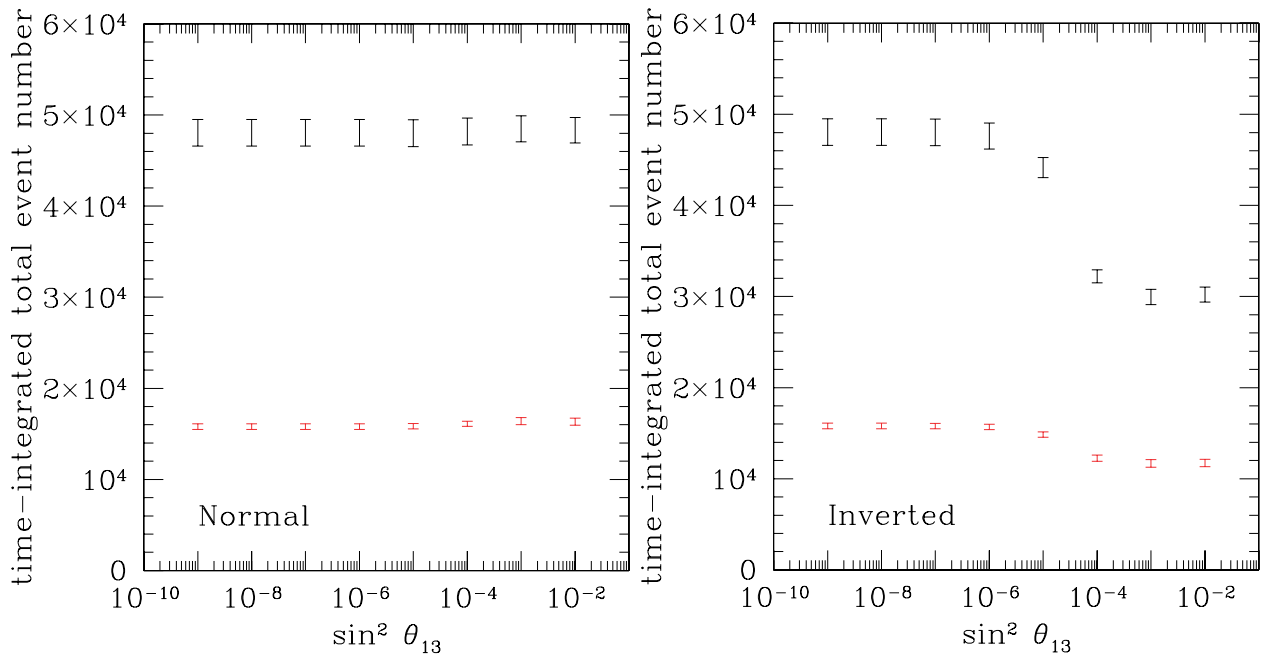}
\caption{Time-integrated total event number of failed supernova neutrinos for the normal mass hierarchy (left) and the inverted mass hierarchy (right). Error bars represents the upper and lower limits owing to the different nadir angles. The upper and lower sets represent models~W40S and W40L, respectively.}
\label{eosdepf}
\end{center}
\end{figure}
As found in Papers~I and II, the luminosity and duration time of failed supernova neutrinos depend on EOS and progenitor model. In Fig.~\ref{eosdepf}, the time-integrated event numbers of models~W40S and W40L are shown for various parameter sets. The total event number of model~W40S is at least 29,124 while that of model~W40L is at most 16,779. This large difference arise from not only its long duration time but also its high average neutrino energy for model~W40S. For all parameter sets, event number of model~W40S is at least 2.5 times larger than that of model~W40L. Estimating from the studies of postbounce evolutions ($<1$~sec) of the neutrino emission \cite{sumi05} and protoneutron star coolings \cite{suzuki06}, the difference by the EOS is not so large for ordinary supernovae. Unfortunately, there is no previous investigation of an EOS dependence of supernova neutrinos from numerical computations of the successful explosion and the successive long-term ($\gtrsim10$~sec after the bounce) emission. However, it is likely that the black hole formation is a better site to probe the EOS of nuclear matter.
\begin{table}[t]
\caption{Time-integrated event number of failed supernova neutrinos for all models with various cases of the parameter sets. Two values of each column represent the upper and lower limits owing to the different nadir angles.}
\begin{center}
\setlength{\tabcolsep}{8pt}
\scalebox{0.75}{
\begin{tabular}{ccccccccc}
\hline\hline
  & Model & W40S & W40L & T50S & Tz50S & T50L & Tz50L & H40L \\
 mass hierarchy & $\sin^2\theta_{13}$ &  &  &  &  &  &  & \\  \hline
 Normal & $10^{-2}$ & 46,922 - 49,727 & 15,958 - 16,726 & 44,766 - 47,077 & 44,766 - 47,077 & 12,579 - 13,140 & 12,579 - 13,140 & 9,485 - 10,115 \\
  & $10^{-3}$ & 47,051 - 49,908 & 15,997 - 16,779 & 44,881 - 47,236 & 44,886 - 47,239 & 12,606 - 13,176 & 12,607 - 13,178 & 9,489 - 10,127 \\
  & $10^{-4}$ & 46,706 - 49,649 & 15,825 - 16,400 & 44,200 - 46,730 & 44,405 - 46,880 & 12,353 - 12,747 & 12,424 - 12,812 & 9,279 - 9,740 \\
  & $10^{-5}$ & 46,529 - 49,473 & 15,552 - 16,124 & 44,051 - 46,606 & 44,040 - 46,596 & 12,227 - 12,632 & 12,231 - 12,636 & 8,960 - 9,184 \\
  & $10^{-6}$ & 46,591 - 49,508 & 15,498 - 16,096 & 44,110 - 46,654 & 44,107 - 46,651 & 12,208 - 12,638 & 12,210 - 12,638 & 8,856 - 9,112 \\
  & $10^{-7}$ & 46,600 - 49,513 & 15,488 - 16,094 & 44,119 - 46,661 & 44,119 - 46,661 & 12,207 - 12,640 & 12,207 - 12,639 & 8,846 - 9,105 \\
  & $10^{-8}$ & 46,601 - 49,514 & 15,487 - 16,093 & 44,120 - 46,662 & 44,120 - 46,662 & 12,207 - 12,640 & 12,207 - 12,640 & 8,845 - 9,104 \\
  & $10^{-9}$ & 46,601 - 49,514 & 15,487 - 16,093 & 44,120 - 46,662 & 44,120 - 46,662 & 12,207 - 12,640 & 12,207 - 12,640 & 8,845 - 9,104 \\
 Inverted & $10^{-2}$ & 29,391 - 31,041 & 11,336 - 12,143 & 29,248 - 30,673 & 29,248 - 30,673 & 9,220 - 9,820 & 9,220 - 9,820 & 7,933 - 8,609 \\
  & $10^{-3}$ & 29,124 - 30,800 & 11,269 - 12,090 & 29,004 - 30,448 & 29,005 - 30,456 & 9,169 - 9,777 & 9,169 - 9,779 & 7,823 - 8,509 \\
  & $10^{-4}$ & 31,502 - 32,945 & 11,913 - 12,600 & 30,717 - 31,762 & 29,197 - 30,245 & 9,335 - 9,710 & 9,119 - 9,554 & 7,599 - 8,116 \\
  & $10^{-5}$ & 43,058 - 45,244 & 14,566 - 15,154 & 40,262 - 42,325 & 38,406 - 40,178 & 11,199 - 11,583 & 10,790 - 11,156 & 8,355 - 8,630 \\
  & $10^{-6}$ & 46,198 - 49,028 & 15,381 - 15,985 & 43,656 - 46,143 & 43,385 - 45,834 & 12,080 - 12,507 & 12,016 - 12,439 & 8,783 - 9,044 \\
  & $10^{-7}$ & 46,560 - 49,465 & 15,476 - 16,082 & 44,073 - 46,609 & 44,045 - 46,577 & 12,194 - 12,626 & 12,187 - 12,619 & 8,838 - 9,098 \\
  & $10^{-8}$ & 46,597 - 49,509 & 15,486 - 16,092 & 44,116 - 46,657 & 44,113 - 46,654 & 12,206 - 12,638 & 12,205 - 12,638 & 8,844 - 9,104 \\
  & $10^{-9}$ & 46,600 - 49,514 & 15,487 - 16,093 & 44,120 - 46,662 & 44,120 - 46,661 & 12,207 - 12,640 & 12,207 - 12,639 & 8,845 - 9,104 \\
\hline\hline
\end{tabular}
}
\label{allresult}
\end{center}
\end{table}

In Table~\ref{allresult}, the summary of event numbers for all models are shown. We can see that the event number of failed supernova neutrinos is mainly determined by properties of the EOS. Especially for the models by Shen-EOS (W40S and T50S), the difference is very minor. As discussed in Paper~II, the duration time of the neutrino emission, which is mainly determined by the EOS, is also affected by the mass accretion rate after the bounce. The accretion rate depends on the density profile of a progenitor and T50S is reported to have lower accretion rate than W40S for late phase. If the accretion rate is lower, the neutrino duration time becomes longer but the neutrino luminosity is lower. This is because the neutrino luminosity is roughly proportional to the accretion luminosity. Therefore both effects are canceled out and the resultant event number becomes similar for models~W40S and T50S. In case of LS-EOS (W40L, T50L and H40L), the event numbers differ by a factor of 1.8 at most and they are roughly proportional to the neutrino duration time. For these cases, the black hole formation occurs fast ($\lesssim 0.6$~s) after the bounce and the difference in neutrino luminosities induced by the accretion rate is not so large for this phase. Differences between the progenitor models are smaller than those between the EOS's for the models investigated in this study.

\begin{figure}
\begin{center}
\includegraphics{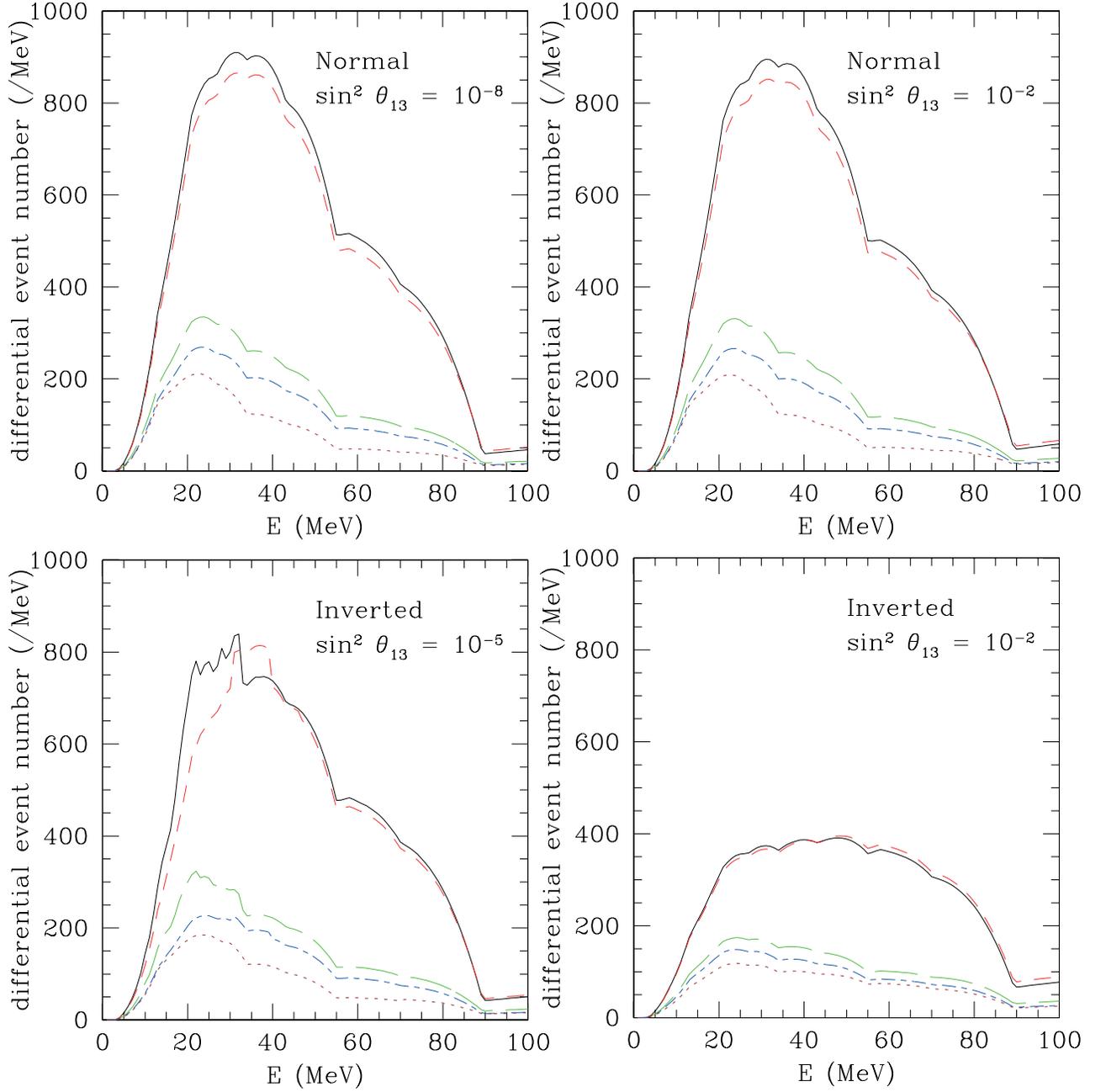}
\caption{Time-integrated spectra for the total event number of failed supernova neutrinos for the normal mass hierarchy with $\sin^2\theta_{13}=10^{-8}$ (upper left), the normal mass hierarchy with $\sin^2\theta_{13}=10^{-2}$ (upper right), the inverted mass hierarchy with $\sin^2\theta_{13}=10^{-5}$ (lower left) and the inverted mass hierarchy with $\sin^2\theta_{13}=10^{-2}$ (lower right). Results obtained without the earth effects are shown. Solid, long-dashed, short-dashed, dot-dashed and dotted lines represent models~W40S, W40L, T50S, T50L and H40L, respectively.}
\label{modedeps}
\end{center}
\end{figure}

We show in Fig.~\ref{modedeps} the spectra of each model for several parameter sets. We can see the clear difference between two EOS's. Moreover, In the case of the inverted mass hierarchy with $\sin^2\theta_{13}=10^{-5}$, another interesting feature is seen in the spectra (lower left panel of Fig.~\ref{modedeps}). It is notable that peaks are mismatched in the comparison between the spectra of models~W40S and T50S. This comes from the different dependence of the neutrino survival probabilities inside the stellar envelope as shown in Fig.~\ref{prog}. The density at H resonance for neutrinos with the energy which gives the peak of spectrum in Fig.~\ref{modedeps} roughly corresponds to that of the boundary of layers, where the density profile becomes steep.

\begin{figure}
\begin{center}
\includegraphics{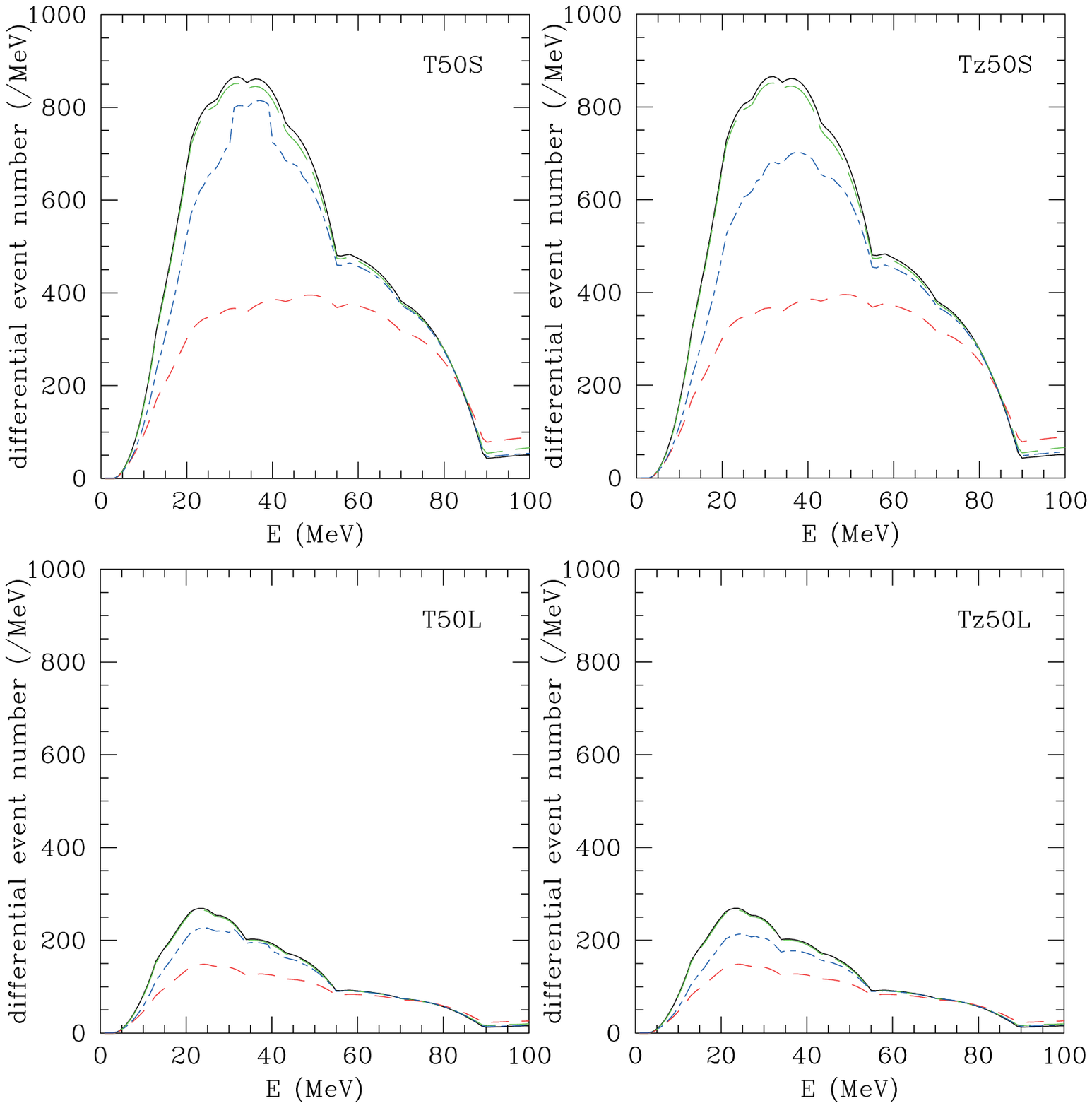}
\caption{Same with Fig.~\ref{paradeps} but for models~T50S (upper left), Tz50S (upper right), T50L (lower left) and Tz50L (lower right) without the earth effects.}
\label{mslsdeps}
\end{center}
\end{figure}

In order to study the impact of the mass-loss of a progenitor, we compare models~T50S, Tz50S, T50L and Tz50L in Fig.~\ref{mslsdeps}. Results for models~T50S and Tz50S are the same except the case of the inverted mass hierarchy with $\sin^2\theta_{13}=10^{-5}$. This is also the case for the comparison of models~T50L and Tz50L. For the regime of a neutrino energy $E=30$-40~MeV, the spectrum of model~T50S has a bump, which does not appear for model~Tz50S. This bump comes again from a steep density profile due to the onion-skin-like structure of the progenitor. The envelope of progenitor model~TUN07Z is stripped by the effects of mass-loss and corresponding boundary is absent. Thus, only when the mass hierarchy is inverted and $\sin^2\theta_{13}\sim10^{-5}$-$10^{-4}$, we have chances of probing the structure of a stellar envelope and effects of mass-loss. Incidentally, mass-loss may affect also the structure of more inner region. If this is the case, the dynamics and neutrino emission themselves may be changed.

In conclusion, the event number of failed supernova neutrinos depends primarily on EOS. Ambiguities on the mixing parameters also affect the result and the spectra of detected neutrinos may be useful for the restrictions of them, while they can also be determined from other experiments. The event number reflects the properties of progenitor models though its dependence is weaker than that of EOS. For the limited case of the mixing parameters, we may be able to probe the structure of a stellar envelope. There is weak dependence on the nadir angle, which should be determined from the identification of the progenitor astronomically. Incidentally, Papers~I and II discussed mainly about the time evolutions and duration times of the neutrino emission. Here, we have focused on the features of time-integrated event numbers and spectra, which are complementary to the previous studies.

We remark that the conclusion given above is based on the results of some restrictive models and more detailed investigations are necessary to examine how general our conclusion is. Incidentally, EOS including the new degree of freedom such as hyperons \cite{takochu} or quarks \cite{self08} are shown to have different features at hot and dense regime recently. A new EOS of the nuclear matter for astrophysical simulations is also proposed to construct based on the many body theory, which is a different approach from Shen-EOS and LS-EOS \cite{kann07}. However, we stress that the ambiguities and model dependences which can be taken into account under the current status have been fully involved and we feel that the essential point has been clarified in our investigations.

Finally, we speculate the observational feasibility. The event number of models computed under Shen-EOS is at least 29,004 while that of LS-EOS is at most 16,779. Let us assume that the signal of failed supernova neutrinos is discovered from the data of the detector and its progenitor is identified in future. These facts indicate that the difference of EOS can be distinguished even if the ambiguities on the mixing parameters and progenitor models are not constrained well. If the mixing parameters are determined till that time by another way, the discrimination of EOS is more promising. In this analysis, the determination of the distance from the progenitor is a key factor. If we are lucky, the progenitor will have already been monitored by the surveys as presented in Ref.~\onlinecite{kocha08}. Even if there are no \textit{ex-ante} surveys, we can determine the direction of the progenitor to some extent by the neutrino detection itself \cite{ando02}. Then the progenitor may be identified from old photographs. In addition, the nadir angle of the progenitor can also be determined so that we can remove the ambiguities of the nadir angle.

The rate of these black hole forming collapses is highly uncertain but will be less than that of ordinary supernovae. Assuming the Salpeter's initial mass function, the black hole formation rate is $\sim25$\% that of ordinary supernovae \cite{kocha08}. Since the supernova rate of our galaxy is $\sim0.03$~/yr \cite{ando05}, we have to wait $\sim130$~years for the black hole formation. One solution to this rate problem is to observe the nearby galaxies by the future detectors with larger fiducial volume. Andromeda galaxy (M31) lies 780~kpc away. In the most optimistic case of Shen-EOS, the event number of neutrinos emitted from it is roughly estimated as $\sim8.2$ for SK. If we consider a future detector with a scale of 5~Mton (e.g. the proposed Deep-TITAND detector \cite{ysuzu01, kist08}), the event number becomes $\sim1900$. This will be enough number to compare the LS-EOS case with $\sim500$ events. Since the cumulative supernova rate within the distance to M31 is $\sim0.08$~/yr \cite{ando05}, we should wait $\sim50$~years. If we extend further out to 4~Mpc, the event number becomes 70 for the most optimistic case. Since the supernova rate within 4~Mpc is $\sim0.3$~/yr, we should wait $\sim13$~years for the black hole formation.

\section{Summary}\label{summary}
In this study, we have evaluated the event number of neutrinos emitted from the stellar collapse involving black hole formation (failed supernova neutrinos) for the currently operating neutrino detector, SuperKamiokande. Since features of emitted neutrinos are affected by the property of equation of state (EOS) and the progenitor model \cite{sumi07, sumi08}, we have investigated the impacts of them under the currently available models. We stress that this is the first study to take into account the effects of flavor mixing for failed supernova neutrinos. Neutrino oscillations have been computed using not only the realistic envelope models of progenitors but also the density profile of the earth. We have also examined the dependence on the mass hierarchy and the mixing angle, $\sin^2\theta_{13}$, in the range allowed up to now.

As a result, we have found that the total event numbers of failed supernova neutrinos are larger than or comparable to those of supernova neutrinos for all cases computed in this study. This result indicates that the black hole formation induced by the stellar collapse is the candidate for the neutrino astronomy as well as the ordinary core-collapse supernovae in spite of the shorter duration time of the neutrino emission. It is also important to note that the failed supernova neutrinos may hardly be negligible to estimate the flux of relic neutrino backgrounds. We have also found that the event number of failed supernova neutrinos depends primarily on EOS. Ambiguities on the mixing parameters also affect the result especially for the case of the inverted mass hierarchy and larger $\sin^2\theta_{13}$. Modifications by the progenitor models are not so drastic comparing with those by EOS, however, the structure of their envelopes affects the result for the limited case, the inverted mass hierarchy and $\sin^2\theta_{13}\sim10^{-5}$-$10^{-4}$. While spectra of events are deformed by the earth effects, their impacts are minor for the total event numbers. These results imply that failed supernova neutrinos are favorable to probe the properties of EOS for hot and/or dense matter.

\begin{acknowledgments}
We are grateful to H. Umeda for providing a progenitor model and fruitful discussions. One of the authors (K. N.) appreciates the cooperation of K. Nakayoshi and Y. Yamamoto received in the early stage of this study. In this work, numerical computations were partially performed on supercomputers at CfCA in the National Astronomical Observatory of Japan, Japan Atomic Energy Agency, YITP in Kyoto University, RCNP in Osaka University and KEK (KEK Supercomputer project 08-13). This work was partially supported by Japan Society for Promotion of Science (JSPS), Grants-in-Aid for the Scientific Research from the Ministry of Education, Science and Culture of Japan through 17540267, 18540291, 18540295, 19104006, 19540252 and the 21st-Century COE Program ``Holistic Research and Education Center for Physics of Self-organization Systems.''
\end{acknowledgments}

\bibliographystyle{apsrev} 
\bibliography{apssamp}% Produces the bibliography via BibTeX.

\end{document}